%% file: PEARC_2024_LLload.tex
\documentclass[sigconf]{acmart}

\usepackage{listings}
\lstdefinestyle{myCustomStyle}{
  frame=single,
  tabsize=4,
  showspaces=false,
  showstringspaces=false
}

\AtBeginDocument{%
  \providecommand\BibTeX{{%
    \normalfont B\kern-0.5em{\scshape i\kern-0.25em b}\kern-0.8em\TeX}}}

\setcopyright{acmlicensed}
\copyrightyear{2024}
\acmYear{2024}
\setcopyright{rightsretained}
\acmConference[PEARC '24]{Practice and Experience in Advanced Research Computing}{July 21--25, 2024}{Providence, RI, USA}
\acmBooktitle{Practice and Experience in Advanced Research Computing (PEARC '24), July 21--25, 2024, Providence, RI, USA}
\acmDOI{10.1145/3626203.3670565}
\acmISBN{979-8-4007-0419-2/24/07}

\begin{document}

\title{LLload: Simplifying Real-Time Job Monitoring for HPC Users}


\author{Chansup Byun}
\authornote{All authors contributed equally to this research.}
\email{cbyun@ll.mit.edu}

\author{Julia Mullen}
\authornotemark[1]
\email{jsm@ll.mit.edu}

\author{Albert Reuther}
\authornotemark[1]
\email{reuther@ll.mit.edu}

\author{William Arcand}
\authornotemark[1]
\email{warcand@ll.mit.edu}

\author{William Bergeron}
\authornotemark[1]
\email{bbergeron@ll.mit.edu}

\author{David Bestor}
\authornotemark[1]
\email{david.bestor@ll.mit.edu}

\author{Daniel Burrill}
\authornotemark[1]
\email{daniel.burrill@ll.mit.edu}

\author{Vijay Gadepally}
\authornotemark[1]
\email{vijayg@ll.mit.edu}

\author{Michael Houle}
\authornotemark[1]
\email{michael.houle@ll.mit.edu}
\affiliation{%
  \institution{MIT Lincoln Laboratory Supercomputing Center}
  \streetaddress{244 Wood Street}
  \city{Lexington}
  \state{MA}
  \country{USA}
  \postcode{02421}
}

\author{Matthew Hubbell}
\authornotemark[1]
\email{mhubbell@ll.mit.edu}

\author{Hayden Jananthan}
\authornotemark[1]
\email{hayden.jananthan@ll.mit.edu}

\author{Michael Jones}
\authornotemark[1]
\email{michael.jones@ll.mit.edu}

\author{Peter Michaleas}
\authornotemark[1]
\email{pmichaleas@ll.mit.edu}

\author{Guillermo Morales}
\authornotemark[1]
\email{guillermo.morales@ll.mit.edu}

\author{Andrew Prout}
\authornotemark[1]
\email{aprout@ll.mit.edu}

\author{Antonio Rosa}
\authornotemark[1]
\email{antonio.rosa@ll.mit.edu}

\author{Charles Yee}
\authornotemark[1]
\email{yee@ll.mit.edu}

\author{Jeremy Kepner}
\authornotemark[1]
\email{kepner@ll.mit.edu}
\orcid{ }
\affiliation{%
  \institution{MIT Lincoln Laboratory Supercomputing Center}
  \streetaddress{244 Wood Street}
  \city{Lexington}
  \state{MA}
  \country{USA}
  \postcode{02421}
}

\author{Lauren Milechin}
\authornotemark[1]
\affiliation{
  \institution{Massachusetts Institute of Technology, Office of Research Computing and Data}
  \city{Cambridge}
  \country{USA}
}
\email{milechin@mit.edu}

\renewcommand{\shortauthors}{Byun, et al.}

\begin{abstract}

One of the more complex tasks for researchers using HPC systems is performance monitoring and tuning of their applications.  Developing a practice of continuous performance improvement, both for speed-up and efficient use of resources is essential to the long term success of both the HPC practitioner and the research project. Profiling tools provide a nice view of the performance of an application but often have a steep learning curve and rarely provide an easy to interpret view of resource utilization. Lower level tools such as \texttt{top} and \texttt{htop} provide a view of resource utilization for those familiar and comfortable with Linux but a barrier for newer HPC practitioners. To expand the existing profiling and job monitoring options, the MIT Lincoln Laboratory Supercomputing Center created \texttt{LLoad}, a tool that captures a snapshot of the resources being used by a job on a per user basis.  \texttt{LLload} is a tool built from standard HPC tools that provides an easy way for a researcher to track resource usage of active jobs. We explain how the tool was designed and implemented and provide insight into how it is used to aid new researchers in developing their performance monitoring skills as well as guide researchers in their resource requests.
\end{abstract}

\ccsdesc[500]{Computing methodologies~Parallel computing methodologies}
\ccsdesc[500]{General and reference~Performance}
\ccsdesc[500]{General and reference~Evaluation}

\keywords{real-time job monitoring, HPC, cluster system, HPC training, HPC tuning}


\maketitle

\input{01_intro_update}

\input{02_design_implementation}

\input{03_training_usage}

\input{05_summary_updated}





\begin{acks}
The authors express their gratitude to Bob Bond, Alan Edelman, Jeffrey Gottschalk, Charles Leiserson, Kristen Malvey, Heidi Perry, Stephen Rejto, Mark Sherman and Marc Zissman for their support of this work. 

DISTRIBUTION STATEMENT A. Approved for public release. Distribution is unlimited.

This material is based upon work supported by the Under Secretary of Defense for Research and Engineering under Air Force Contract No. FA8702-15-D-0001. Any opinions, findings, conclusions or recommendations expressed in this material are those of the author(s) and do not necessarily reflect the views of the Under Secretary of Defense for Research and Engineering.

\copyright 2024 Massachusetts Institute of Technology.

Delivered to the U.S. Government with Unlimited Rights, as defined in DFARS Part 252.227-7013 or 7014 (Feb 2014). Notwithstanding any copyright notice, U.S. Government rights in this work are defined by DFARS 252.227-7013 or DFARS 252.227-7014 as detailed above. Use of this work other than as specifically authorized by the U.S. Government may violate any copyrights that exist in this work.
 
\end{acks}

\bibliographystyle{ACM-Reference-Format}
\bibliography{PEARC_2024_LLload}

\end{document}

%% file: 01_intro_update.tex
\section{Introduction}

One of the more difficult aspects of High Performance Computing (HPC) training is guiding practitioners as they learn how to request resources appropriate to their application.  Correctly aligning resource requests with application needs requires understanding the memory and compute needs for a given software application and its data. On one end of the spectrum, requesting too few compute resources can lead to excessive run-times, \textit{which could impact queue lengths and times}.  More significantly, insufficient memory resources can lead to out-of-memory errors and node instabilities that directly impact other user applications running on the node. To avoid these issues, many HPC centers have established guardrails to insure that a single user application cannot negatively impact other users. At the other end of the spectrum are users whose resource requests overwhelm their actual need, resulting in wasted resources.  To our knowledge there are no easy-to-use, community-wide tools to address this under-utilization of resources.

 Two key challenges for HPC trainers include the "more is better" user perspective and the bespoke nature of user jobs, which means that there is no "one size fits most" method for calculating resource requirements.  The best that trainers can do is to provide a set of guidelines for evaluating an application and determining the resource requirements. These guidelines outline a process by which users can profile and monitor code at run-time, using a range of high and low level tools and profilers, for example~\cite{utah,hpc-wiki,nersc}.  Over the decades, the HPC community has created many powerful profiling tools, some have steep learning curves ~\cite{tau}, while others require machine level counters that can affect run-time or require administrator privileges for installation, e.g., Intel Vtune Profiler, Linux perf. Standard Linux tools such as \texttt{top} and \texttt{htop} provide a means for viewing the workloads on a node, while vendor tools such as \texttt{nvidia-smi}, \texttt{rocm-smi}, and \texttt{nvtop} offer insight into GPU utilization. Further, XD-Mod~\cite{xdmod} and TACC Stats~\cite{taccstats} can be used to collect job execution data that can be made available to users for analysis after their job has completed. Ganglia~\cite{ganglia} collects a wide ensemble of execution data and can display it to users during execution, but the variety of the ensemble of metrics can quickly overwhelm most novice users. For applications that have a long shelf-life, or require real-time performance, using these tools effectively can highlight areas in need of tuning.  
 
 However, researchers using packaged libraries to proto-type devices, workflows and algorithms often have limited means of tuning the software and often find the overhead of learning profiling tools too high a barrier for short-term mission-driven applications.  To address this, the team at the MIT Lincoln Laboratory Supercomputing Center (LLSC) created \texttt{LLload} a tool that captures a snapshot of the resources being used by a job on a user basis.  The \texttt{LLload} tool provides an easy way for the user to track their resource usage, for Operations Team to monitor under-utilization of resources during busy times and simplifies the training involved in helping users understand how to determine their resource requirements.

%% file: 02_design_implementation.tex
\section{LLload}

\subsection{Design}
The LLSC has a long history of lowering the barrier to entry for supercomputing practitioners by creating an environment where researchers can focus on short term mission driven applications~\cite{lessons-learned-llsc}.  As it became apparent that researchers needed a simplified, intuitive approach for job monitoring, the team began to design a tool that would
\begin{itemize}
    \item simplify the monitoring of active jobs,
    \item track jobs in real-time, 
    \item display information about CPU, GPU and memory usage, and
    \item display information in a clear, human readable manner,
\end{itemize}
in order to support users as they developed job-monitoring and profiling skills.  To keep the tool from impacting the system, especially with a large group of active users, the tool needed to 
\begin{itemize}
    \item be light-weight,
    \item not interfere with operation of the cluster system even under heavy user load, and
    \item use existing tools where possible.
\end{itemize}
Further, because it is quite challenging to collect and attribute process execution metrics to the appropriate users' processes when multiple users have jobs executing on the same compute node, we needed to configure the scheduler for per-node, single-user scheduling.

\subsection{Implementation}

Providing and displaying information about CPU, GPU and memory use, while the job is running, requires first collecting the appropriate data. For HPC systems running a scheduler, most of these data can be collected via the scheduler.  The LLSC systems use SLURM as the scheduler and the \texttt{sinfo}, \texttt{scontrol}, and \texttt{squeue} commands will capture the list of active users, CPU loads and memory usage. Retrieving information about the GPU loads requires a vendor utility since SLURM does not include information on GPU loads.  The LLSC clusters have NVIDIA GPUS and \texttt{nvidia-smi} is used to collect the GPU data. Since the command \texttt{nvidia-smi} needs to be executed on the GPUs' host node, it is important to maintain passwordless ssh connection capabilities to all compute nodes in the system so that if the GPU information is needed the appropriate ssh remote connections can be performed on the GPUs of interest.

\begin{figure}[htbp]
   \centering
   \includegraphics[width=3.3in]{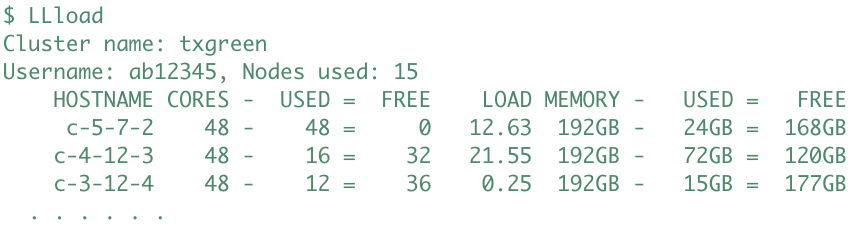}
  \caption{The default behavior of \texttt{LLload}.}
  \label{LLloadDefault}
  \Description{A typical output showing the system usage on a node with \texttt{LLload}.}
\end{figure}
\begin{figure}[htbp]
   \centering
   \includegraphics[width=\linewidth]{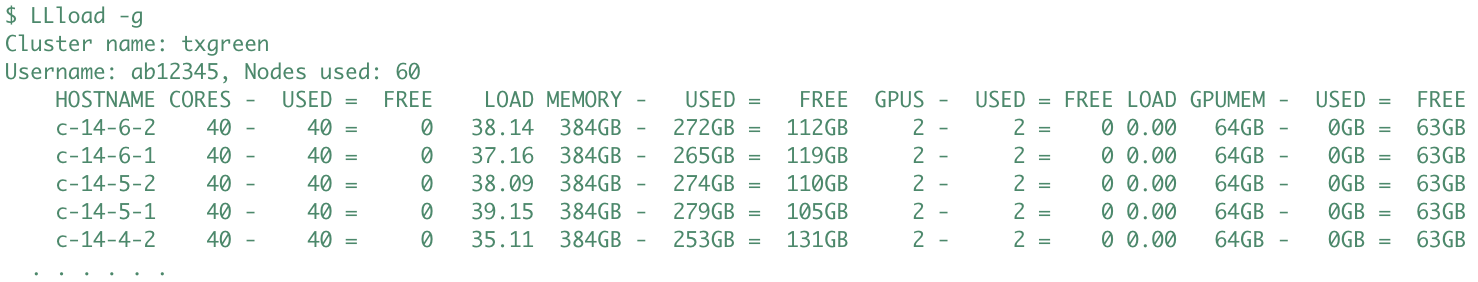}
  \caption{Typical output of \texttt{LLload} with the \texttt{-g} option.}
  \label{LLloadGPU}
  \Description{An output showing the system usage along with GPU usage on a node using \texttt{LLload}.}
\end{figure}

When a user executes the \texttt{LLoad} command, the following steps are taking to collect and display their jobs' execution information: 
\begin{enumerate}
    \item execute the SLURM command \texttt{sinfo} to identify all of the nodes assigned for Jupyter notebook jobs 
    \item execute the SLURM command \texttt{squeue} to identify all of the active jobs and nodes on which the jobs are executing. 
    \item execute the SLURM command \texttt{sinfo} to extract the details of system status of all nodes on which the user's jobs are running. 
    \item If the applications use GPUs, 
    perform a remote ssh execution of \texttt{nvidia-smi} to get the GPU usage information on each individual node. The information about GPU usage includes the number of available and used GPUs, the GPU load, and GPU memory usage. 
    \item populate an output template with the appropriate collected data and print it to the screen.
\end{enumerate}
In the first four steps, the command is called, the command text response is captured and parsed for pertinent information, and a Python dictionary is populated with this data. The information in that Python dictionary is then used to populate an output template to print out the results to the screen in step five. It should be noted that the command output parsing can be extensive, depending on the formatting and content of the command output. For example, the \texttt{squeue} command is used to identify all of the active jobs and nodes on which the jobs are executing. Using command line arguments, the username is specified so that we do not get a list of all users' jobs that are running, and the output format can be customized (e.g., comma-delimited items of only the needed job metadata fields). With that output, there is still parsing involved to make sure appropriate data has been returned and that the Python dictionary is populated with the appropriate data items.
The output from both the basic \texttt{LLoad} and GPU specific \texttt{LLload -g} commands are shown in Figure~\ref{LLloadDefault} and Figure~\ref{LLloadGPU} respectively.


While implementing the LLload design, a few challenges surfaced. One is that the SLURM command {\tt sinfo} does not enable finer-grained filtering based on usernames or nodelists; it outputs all jobs that are currently executing, regardless of user. We would preferred to only receive as output information about the nodes on which the jobs of the user is executing. Instead, the output needs to be filtered against a list of nodes on which an individual user's jobs is executing before we extract the pertinent information and populate the Python dictionary. 

Another challenge was collecting GPU usage information. We would have preferred to extract this information from Slurm, but this functionality is not yet available.   Most of other HPC centers seems to be using either the GPU vendors commands directly or a separate server monitoring GPU usage.  The latter approach seems to provide a lot more capability but it is difficulty to set up.  Hence, we decided to use the GPU vendor command to extract the GPU usage which only provides a snapshot of GPU utilization, GPU memory usage, power consumption, etc. at the time when the query is made. That is, the result may be misleading because the data is not collected over a time window and statistically process as the CPU load metrics are in Linux. In order to get mre meaningful GPU metrics, one would need to run the GPU metric command multiple times and compute our own statistics, but this is not currently implemented in \texttt{LLload}. Naturally, users could execute \texttt{LLload} multiple times in fairly rapid succession in order to check if the usage information varies significantly.



%% file: 03_training_usage.tex
\section{Integrating LLload into Research Facilitation}

It is standard practice for HPC practitioners to learn how to analyze application performance in order to improve speed up, insure that there are no out-of-memory issues and understand the application resource requirements.  For serial applications, there are a range of profilers, most language and library specific, to evaluate the application performance, identify sections of an application, model or code in need to optimizing, and report memory use. The information returned from a profiler is key to improving performance of the application, but lacks information about the resource usage and how well that usage aligned with the resource request. One approach to evaluating application resource usage, particularly for high-throughput applications, involves using standard application analysis tools such as \texttt{htop}, \texttt{top}, and \texttt{nvidia-smi}. For researchers who are new to HPC, this can be a heavy lift and \texttt{LLload} was created to lower this barrier to entry.  In order to truly simplify the process however, researchers need to understand how to interpret the results from the tools and how to use them to modify resource requests.  

The development of new training materials followed the normal process whereby members of the research facilitation team augmented the job monitoring documentation~\cite{supercloud-llload} to include \texttt{LLload}, while at the same incorporating the information into direct consultations with researchers.  The documentation explains that the LLload command provides information to help you  evaluate the efficiency of your jobs and lists the information returned by the \texttt{LLLoad} command;  all of the nodes that you have jobs running on, how many of the cores on those nodes you have allocated, and some statistics about how the resources on those nodes are being used, including the 5 minute average CPU load, the memory utilization plus the memory used for caching, the GPU utilization and GPU memory.  While this provides the researcher with a profile of their usage, what the researcher really needs is target values and guidance on how to reach them.  To aid the researcher, documentation and direct consultation recommend a target for CPU usage is 50-150 \% of the number of CPUs on the node. Furthermore, the researcher is informed that if this number is lower, it's likely they can take advantage of more resources on the node. Alternatively, if they find the load numbers are very high, they are warned that they risk speed-down or even overwhelming the node. Finally, the researcher is advised that they  have a few knobs to turn to adjust CPU utilization, for example, changing the number of threads used by the application or by running more jobs or processes per node.  
    
With respect to memory usage, the returned value is a composite, so the researcher is advised to \texttt{ssh} to the node and run \texttt{htop} to see the true memory utilization, or use the \texttt{sacct} command after the job has completed to get the peak memory utilization. Additional documentation is provided for using these commands. For researchers who are using GPUs, the research facilitator explains that the GPU load is normalized such that 100 \% utilization on both GPUs will give a value of 2, so if both GPUs are well utilized they should expect to see a value close to 2.  The displayed GPU values are derived from an instantaneous value rather than averaged over a period of time and researchers are told to run the \texttt{LLload} command a few times to get a good idea of GPU utilization.  Researchers with low GPU utilization are directed to documentation about optimizing GPU usage.

 In order to start new users off on the correct path, the LLSC-SuperCloud team provides learning modules on ``Best Practices in Scalable Development'' within our Practical HPC course~\cite{PEARC-PHPC}.The module provides a stepwise approach to develop concurrent code from serial applications, explains how to analyze the serial application in advance of scaling and includes hands-on training with many of standard application analysis tools mentioned in the introduction.  After vetting the \texttt{LLload }documentation and gathering feedback from direct consultations, the \texttt{LLload} command has been integrated into workshops and the online course.  

%% file: 05_summary_updated.tex
\section{Summary}
One of the more complex tasks for researchers using HPC systems is performance monitoring and tuning of their applications.  Developing a practice of continuous performance improvement, both for speed-up and efficient use of resources is essential to the long term success of both the HPC practitioner and the research project. To lower the complexity of job monitoring, we have introduced \texttt{LLload}, a real-time job monitoring tool that provides researchers with an easy way to evaluate the efficiency of their jobs. \texttt{LLload} works at the command line of a login or compute node and returns a snapshot of the CPU, GPU, and memory resources associated with the researchers active jobs. To  increase effective use of the tool,  the research facilitation team has developed documentation, examples and online course materials to help the researcher interpret the \texttt{LLload} results and modify their process to improve efficiency and performance. In closing, we note that the \texttt{LLload} tool is build using standard HPC scheduler and GPU command and thus can be built for any HPC or Research Computing and Data center.